\makeatletter

\newcommand{\Rmnum}[1]{\expandafter\@slowromancap\romannumeral #1@}
\makeatother

\def\kms{$\rm km\,s^{-1}$} 

\def\Msun{${\rm M}_\odot$}
\def\Msunyr{${\rm M}_\odot\,{\rm yr}^{-1}$}
\def\thco{$^{13}$CO}
 
\def\twco{$^{12}$CO}
\def\twcsio{$^{12}$C$^{16}$O}
\def\ceo{C$^{18}$O}

\def\cth{$^{13}$C}

\def\oei{$^{18}$O}

\def\sig{$\sigma$}
\def\j{$J$}
\def\to{$\rightarrow$}
\def\Lir{$L_{\rm IR}$}
\def\target{SDSS\,J0901+1814}
\def\z{{\it z}}

\documentclass[twocolumn]{aastex631}
\usepackage{enumitem}
\usepackage{amssymb}
\usepackage{soul}

\shorttitle{}
\shortauthors{Guo et al.}
\graphicspath{{./}{figures/}}

\begin{document}

\def\purple#1 {{\textcolor{purple}{#1}}\ }
\def\red#1 {\textcolor{red}{#1}}
\def\blue#1 {{\textcolor{blue}{#1}}\ }
\definecolor{forestgreen}{rgb}{0.13, 0.55, 0.13}
\def\zy#1 {\textcolor{forestgreen}{ zy: #1}}
\def\ziyi#1 {\textcolor{purple}{ ziyi: #1}}
\newcommand{\xtnote}[1]{\textit{\textcolor{orange}{[xiaoting:#1]}}}
\newcommand{\xt}[1]{\textcolor{orange}{#1}}
\newcommand{\egnote}[1]{\textit{\textcolor{magenta}{[Eda: #1]}}}
\newcommand{\eg}[1]{\textcolor{magenta}{#1}}

\title{First detection of CO isotopologues in a high-redshift main-sequence galaxy: evidence of a top-heavy stellar initial mass function}

\correspondingauthor{Zhi-Yu Zhang}
\email{zzhang@nju.edu.cn}

\author[0000-0002-7532-1496]{Ziyi Guo}
\affiliation{School of Astronomy and Space Science, Nanjing University, Nanjing 210093, China}
\affiliation{Key Laboratory of Modern Astronomy and Astrophysics (Nanjing University), Ministry of Education, Nanjing 210093, China}
\author[0000-0002-7299-2876]{Zhi-Yu Zhang}
\affiliation{School of Astronomy and Space Science, Nanjing University, Nanjing 210093, China}
\affiliation{Key Laboratory of Modern Astronomy and Astrophysics (Nanjing University), Ministry of Education, Nanjing 210093, China}
\author[0000-0001-7395-1198]{Zhiqiang Yan}
\affiliation{School of Astronomy and Space Science, Nanjing University, Nanjing 210093, China}
\affiliation{Key Laboratory of Modern Astronomy and Astrophysics (Nanjing University), Ministry of Education, Nanjing 210093, China}
\author[0000-0002-7440-1080]{Eda Gjergo}
\affiliation{School of Astronomy and Space Science, Nanjing University, Nanjing 210093, China}
\affiliation{Key Laboratory of Modern Astronomy and Astrophysics (Nanjing University), Ministry of Education, Nanjing 210093, China}
\author[0000-0003-2475-124X]{Allison Man}
\affiliation{Department of Physics and Astronomy, University of British Columbia, Vancouver, British Columbia, Canada}
\author[0000-0001-5118-1313]{R.\,J.~Ivison}
\affiliation{European Southern Observatory, Karl-Schwarzschild-Strasse 2, D-85748 Garching bei München, Germany}
\affiliation{Dublin Institute for Advanced Studies, Dublin, Ireland}
\affiliation{Centre of Excellence for All Sky Astrophysics in 3 Dimensions (ASTRO 3D), Australia}
\affiliation{Institute for Astronomy, University of Edinburgh, Blackford Hill, Edinburgh EH9 3HJ, UK}
\author[0000-0002-6506-1985]{Xiaoting Fu}
\affiliation{Purple Mountain Observatory, Chinese Academy of Sciences, Nanjing, China}
\author[0000-0002-8614-6275]{Yong Shi}
\affiliation{School of Astronomy and Space Science, Nanjing University, Nanjing 210093, China}
\affiliation{Key Laboratory of Modern Astronomy and Astrophysics (Nanjing University), Ministry of Education, Nanjing 210093, 
China}



\begin{abstract}

Recent observations and theories have presented a strong challenge to the
universality of the stellar initial mass function (IMF) in extreme environments.
A notable example has been found for starburst conditions, where evidence
favours a top-heavy IMF, i.e.\ there is a bias toward massive stars compared to
the IMF that is responsible for the stellar mass function and elemental abundances
observed in the Milky Way. Local starburst galaxies have star-formation rates
similar to those in high-redshift main-sequence galaxies, which appear to
dominate the stellar mass budget at early epochs. However, the IMF of
high-redshift main-sequence galaxies is yet to be probed. Since \thco\ and
\ceo\ isotopologues are sensitive to the IMF, we have observed these lines
towards four strongly-lensed high-redshift main-sequence galaxies using the
Atacama Large Millimeter/sub-millimeter Array. Of our four targets, \target, at
$z \approx 2.26$, is seen clearly in \thco\ and \ceo, the first detection of
CO isotopologues in the high-redshift main-sequence galaxy population. The
observed \cth/\oei\ ratio, $2.4 \pm 0.8$, is significantly lower than that of
local main-sequence galaxies. We estimate the isotope ratio, oxygen abundance
and stellar mass using a series of chemical evolution models with varying
star-formation histories and IMFs. All models favour an IMF that is more
top-heavy than that of the Milky Way. Thus, as with starburst galaxies,
main-sequence galaxies in the high-redshift Universe have a greater fraction of
massive stars than a Milky-Way IMF would imply. 

\end{abstract}



\section{Introduction}
\label{section: introduction}

When interpreting observations of starlight, infrared (IR) or radio continuum
emission from galaxies -- all typically dominated by energy from massive stars
-- the choice of the stellar initial mass function (IMF) is one of the most
important assumptions we make. This is true whether one is looking at
individual galaxies, or averaging over cosmological volumes
\citep{kennicutt1998b, madau2014}. The choice of IMF has profound implications
for the determination of star-formation rates (SFRs) and stellar masses, for
example, both of which are used to define the so-called `main sequence' of
galaxies. The IMF can thus play a fundamental role in cosmological simulations
of structure formation \citep{baugh2005}, and in the interpretation of galaxy
evolution across cosmic time \citep{daddi2007}. 

The IMF is usually assumed to be a universal and invariant function, based on
observations in the Galaxy and the Magellanic Clouds \citep{kroupa2001,
Bastian2010}. Its universality can be understood in terms of the initial
conditions for star formation, which occur in dense region deep inside
molecular clouds and is well shielded from ultraviolet radiation and where
supersonic turbulence is fully dissipated, so thermodynamically similar
\citep{elmegreen2008}. Cosmic rays (CRs) seem to be the only form of feedback
that can strongly perturb the otherwise well-shielded initial conditions of
star formation. Once a CR energy density background of $\sim $$10^2\times$
Galactic average has been reached \citep[the Galactic average CR energy density
is around 0.5--1.4 eV cm$^{-3}$; ][]{yoasthull2016,papadopoulos2011}, as
expected in starburst environments, the IMF may then be quite different from
that of ordinary star-forming disks \citep{papadopoulos2011}.

An increasing body of evidence points to a non-universal IMF, supported by
observations of stellar properties in external galaxies
\citep{Rieke1993,lee2009,gunawardhana2011,yan2020,mucciarelli2021}, CO isotope
abundances in the interstellar medium \citep[ISM; ][]{romano2017, zhang2018},
and star counting in both the Milky Way and the Large Magellanic Cloud
\citep{Geha2013,li2023,Schneider2018}. It has been proposed that a
non-universal IMF can account for the properties of galaxies under different
physical conditions and at different evolutionary stages
\citep{Matteucci1990,yan2017,jerabkova2018,Hopkins2018,Smith2020}. 

In dust-obscured, gas-rich star-forming galaxies there is no hope of following
the classical method of probing their IMF directly, based on an assessment of
the stellar light via optical/near-IR observations. However,
sub-millimeter/millimeter (sub-mm/mm) wavelengths are largely unobscured by
dust and provide an alternative probe of the IMF, opening a window to measure
the isotopic abundances of particular elements in these systems. The isotopes
of carbon, nitrogen and oxygen \citep[see][for a recent review]{romano2022} are
particularly suitable for this purpose, since they are produced through
different stellar nucleosynthesis processes in stars of different masses, and
are thus released to the ISM on different timescales \citep{kobayashi2011,
romano2017, zhang2018}. For instance, \oei\ is produced predominantly by
massive stars ($\geq 8$\,\Msun) purely through secondary channels, while \cth\
is produced mainly by low- and intermediate-mass stars (LIMS, $< 8$\,\Msun),
through primary plus secondary element channels
\citep[e.g.,][]{portinari1998,marigo2001,romano2017}\footnote{In metal-poor
systems, efficient stellar mixing induced by rotation in low-metallicity,
massive stars may change this picture \citep[e.g.][]{romano2019}.}.

These isotopes are ejected into the ISM via stellar winds \citep{romano2022},
where they form molecules in the same way as their major isotopes, with similar
astrochemical properties. Emission from \thco\ and \ceo\ transitions
(isotopologues of \twcsio) can therefore be used to trace the \cth\ and \oei\
abundances produced by ancestral generations of stars. Abundance ratios of
\thco/\ceo\ have been found to anti-correlate with the IR luminosity of large
molecular gas reservoirs in local galaxies \citep{jimenezdonaire2017,
zhang2018}. The line ratio is found to be roughly unity in ultraluminous IR
galaxies (ULIRGs) \citep{henkel2014, sliwa2017, brown2019}, in contrast to
values ranging from 7 to 10 in the disk of the Milky Way and in local disk
galaxies \citep{giannetti2014, jimenezdonaire2017}. This can be and has been
interpreted as an indication of a higher fraction of massive stars (i.e.\ a
top-heavy IMF) in high-SFR galaxies.

Strengthening the correlation between IR luminosity (\Lir) and the \thco/\ceo\
ratio found in local galaxies, low \thco/\ceo\ ratios have been found in
strongly lensed starbursts at $z\sim 2$--3 \citep{danielson2013, yang2023},
which can be well reproduced in galactic chemical evolution (GCE) models with a
top-heavy IMF \citep{zhang2018}. 

However, it is unclear whether high-redshift main-sequence galaxies -- thought
to contribute the majority of the cosmic SFR \citep{daddi2007, madau2014} --
still hold to the IMF adopted in most current work \citep[e.g.,][]{huang2023}.
In fact, these galaxies are known to have 1--2 orders of magnitude higher SFRs
\citep{schreiber2015} and smaller sizes \citep[and, therefore, higher SFR
surface densities;][]{conselice2014}, compared to those of local main-sequence
galaxies with the same stellar masses. If these high-redshift main-sequence
galaxies were in the local Universe, they would be classified as starbursts
\citep{schreiber2015}, raising the possibility that their star formation also
proceeds with a top-heavy IMF. 

As with their dust-enshrouded starburst brethren, it is difficult to examine
the intrinsic stellar light -- the ultraviolet/optical spectral energy
distribution (SED) -- of high-redshift main-sequence galaxies, because they
contain high fractions of gas and dust \citep{vandergiessen2022}. This prevents
the use of classical methods to derive their IMFs. However, recent Atacama
Large Millimeter/sub-millimeter Array (ALMA) observations have detected CO
emission from a few strongly-lensed, high-redshift main-sequence galaxies
\citep[e.g.][]{dessauges2015,dessauges2017,sharon2019}, which may allow the
application of the \cth/\oei\ diagnostic, despite the weak emission expected
from these isotopologues. 

Isotopic abundances are sensitive to the IMF, to the star-formation history
(SFH), and to stellar nucleosynthesis, thus interpreting them requires
sophisticated GCE modelling \citep{matteucci2001}. As with previous work on
starburst galaxies \citep{romano2017, zhang2018}, we use literature estimates
of the stellar mass and metallicity ([O/H]) of the systems to constrain our GCE
models. For this work, we use a modified version of the open-source GCE code,
NuPyCEE \citep{cote2017, ritter2018}, to simulate the evolution of the
galaxies.

This paper is organised as follows: in Section~\ref{section: observation}, we
describe our observations, data reduction, and the ancillary data adopted in
this work. Section~\ref{section: results} presents the observational results.
In Section~\ref{section: GCEmodel}, we describe the assumptions and settings in
the GCE model. Section~\ref{section: comparison} presents a comparison of the
observational and modelling results. Section~\ref{section: discussion}
discusses the possible IMF bias from SFH and spatial variation and the impact
on the history of cosmic star formation. Throughout this paper, we use the
\citet{planck2016} cosmology, with $\Omega_{\rm m}$ = 0.315, $\Omega_\Lambda$ =
0.685, $\Omega_{\rm b}$ = 0.0490, $h$ = 0.6731, $\sigma_8$ = 0.829 and $n_{\rm
S}$ = 0.9655.


\section{Observations and data reduction}
\label{section: observation}

\subsection{ALMA observations} 

We have gathered a sample of four strongly lensed star-forming galaxies from
the literature. Their peak flux density in a CO line is at least 10\,mJy (cf.\
Table~\ref{table: physical_data}). Their range of redshifts spans $z=1.5$--3.6,
which corresponds to roughly the epoch of maximal cosmic SFR density
\citep[e.g.][]{madau2014}. 

\begin{figure} \centering
        \includegraphics*[width=\columnwidth]{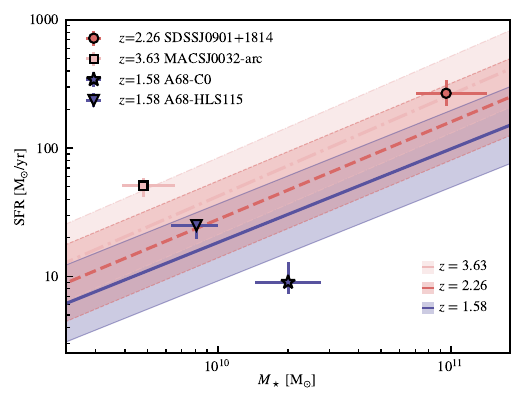}
        \caption{\label{fig: main_sequence} Bands of SFR as a function of
        stellar mass at three different redshifts for main-sequence galaxies
        \citep[][shaded areas represent $\pm 1~\sigma$]{schreiber2015}. The
        symbols refer to our target (main-sequence) galaxies.} 
\end{figure}

The intrinsic (i.e.\ corrected for lensing) SFRs and stellar masses of these
galaxies are in the range 25--600\,\Msunyr\ and 5--$300 \times 10^9$\,\Msun,
respectively \citep{dessauges2017, dessauges2015, sharon2019}. 
The SFR has been estimated from both the H$\alpha$ and from the total IR
luminosity \citep{sharon2019}; we adopt the latter because the SFR estimated
from total IR luminosity is not affected by the dust obscuration in contrast to
H$\alpha$. For the other three galaxies in our sample, the SFRs were
determined from the best SED fits to the combination of the ultraviolet and IR
luminosities. Fig.~\ref{fig: main_sequence} presents all targets superimposed
on the corresponding three main-sequence scaling relations, for $z=1.58$ (two
target galaxies), 2.26 and 3.63; different colors refer to the different
redshifts \citep[see e.g.][]{schreiber2015}. Our four targets are all located
in or under their corresponding main sequence. The oxygen abundance of \target\
is derived from the [N\,{\sc ii}] and H$\alpha$ emission lines
\citep{sharon2019, pettini2004}, while for MACS J0032-arc it is estimated via
the mass-metallicity relation \citep{dessauges2017, pettini2004}.

We observed the \thco\ and \ceo\ transitions towards our target galaxies using
ALMA (programme~2018.1.00588.S: P.I.~Zhi-Yu~Zhang). The full 12-m array was
employed, with 43--52 antennas. MACS\,J0032-arc, A68-CO and A68-LHS115 were
observed in band 4; \target\ was observed in band 3. Before and after each
10-min interval spent on the targets, we switched to a calibrator to track the
complex gain fluctuations. For each spectral window (spw), we recorded 1920
channels, with a spectral resolution of 976.562\,kHz ($\sim $1--3\,\kms). For
each spw, the total bandwidth is 1.875\,GHz. We use one spw to cover both
\thco\ and \ceo, which should minimise any systematic effects between spws.
Other spws were configured in the continuum mode. The coordinates of the
pointing centres and the central observed frequencies are listed in
Table~\ref{table: date}. The total on-source integration time of the project
was 6.93\,hr. Each target had an on-source time of 1--2\,hr. The precipitable
water vapour (PWV) levels were below 6\,mm for all observations. The detailed
observational information, including the choice of calibrators, is listed in
Table~\ref{table: date}.

\subsection{Data reduction } 

We used the common astronomy software package (CASA) \citep{mcmullin2007} to
reduce the data. After the standard pipeline calibration, we fitted the
continuum using line-free channels in two adjacent spws, using the task
`uvcontsub'. For the velocity range of the emission lines we adopted the \twco\
velocity range from the literature (cf.\ Table~\ref{table: physical_data}). We
imaged (with CASA version 6.1.2.7) the visibility data using the `tclean' task
with natural weighting, for both continuum and line data. For the line data, we
set a channel width of 50--80\,\kms. For both continuum and line data, we
adopted an outer uvtaper of 100 k$\lambda$, to maximise sensitivity. The noise
levels in the continuum maps and spectra (see Table~\ref{table: rms}) are
better than 0.015\,mJy\,beam$^{-1}$ and 0.13\,Jy\,km\,s$^{-1}$, respectively.

\subsection{Archival data}
\label{section: archival}

We retrieved archival optical and near-IR images from the {\it Hubble Space
Telescope (HST)} \footnote{\url{https://hla.stsci.edu/}} to help determine the
target positions and compare them with our ALMA data. For \target\ and A68~C0,
we obtained images from Wide Field Camera 3 (WFC3) through the F475W and F110W
filters, respectively. For MACS\,J0032-arc and A68~HLS115, we adopted images
observed through the F814W filter with Wide Field Camera 1 (WFC1). All the HST
data we used in this paper were obtained from the Mikulski Archive for Space
Telescopes (MAST) at the Space Telescope Science Institute. The specific
observations analyzed can be accessed via
\dataset[DOI:10.17909/hqpv-7w15]{https://doi.org/10.17909/hqpv-7w15}. 

We also retrieved the \twco\ $J=3$--2 data (programme~2016.1.00406.S:
P.I.~Dieter~Lutz) for \target\ from the ALMA archive. We used the standard CASA
pipeline to calibrate the data, and then imaged it with a restoring beamsize of
3$''$ and a velocity resolution of 15\,\kms, the same as our \thco\ and \ceo\
data. 

\begin{table*}
    \centering
    \caption{\label{table: physical_data} Information about the targeted
        galaxies from the literature. All information is corrected for lensing
        magnification.}

    \begin{tabular}{llllllll}
        \hline  \hline
        Target          & \z       & \twco\      & \twco\ flux    & 12+log(O/H) & SFR               & $M_{\star}$         & Reference\\
        galaxy          &          & Transition  & [Jy$\cdot$km s$^{-1}$]  &             & [\Msunyr]           & 10$^9$\,\Msun         & \\
        \hline
        \target         & 2.2597   & 3--2        & 19.8$\pm$2.0   & 8.7$\pm$0.2 & 268$^{+63}_{-61}$ & 95$^{+38}_{-28}$    &\citet{sharon2019}\\
        MACS\,J0032-arc & 3.6314   & 6--5        & 2.8$\pm$0.6    & 8.0$\pm$0.2 & 51$^{+7}_{-10}$   & 4.8$^{+1.5}_{-1.0}$ &\citet{dessauges2017}\\
        A68-HLS115      & 1.5854   & 2--1        & 2.0$\pm$0.3    & --            & 25$^{+2.6}_{-6}$  & 8.1$^{+1.7}_{-2.0}$ &\citet{dessauges2015}\\
        A68-C0          & 1.5859   & 2--1        & 1.9$\pm$0.3    & --            & 9$^{+3.3}_{-1.9}$ & 20$^{+6.5}_{-6.5}$          &\citet{dessauges2015}\\
        \hline
    \end{tabular} \\
\end{table*}

\begin{table*}
    \centering
    \caption{Observational information. \textbf{Note:} All the frequency bands
    are included in every observation for the same target.} 
    \footnotesize
    \begin{tabular}{llllcclcc}
         \hline \hline
         Target         & R.A.         & Dec.           & Freq. Range    & Date        & Time$^{\rm obs}$  & Gain cal.& Flux/BP cal. & PWV$^{\rm mean}$ \\
         galaxy         & (J2000)      & (J2000)        & [GHz]          &             & [hr]  &            &            & [mm]      \\
         \hline                                                                                             
        \target         & 09:01:22.4   & +18:14:30     & 98.4--102.2  & 2019-Jan-27 & 0.83 & J0854+2006 & J0750+1231 & 8.8      \\
                         &              &               & 110.6--114.4 & 2019-Mar-27 & 0.83 & J0908+1609 & J0750+1231 & 3.7      \\
                         &              &               &                & 2019-Mar-27 & 0.85 & J0908+1609 & J0725$-$0054 & 3.7      \\
         MACS J0032-arc  & 00:32:07.776 & +18:06:47.80   & 139.7--143.4 & 2018-Dec-20 & 1.10 & J0019+2021 & J0006$-$0623 & 2.0      \\
                         &              &               & 152.0--155.4 & 2019-Jan-24 & 1.13 & J0019+2021 & J0006$-$0623 & 7.1      \\
                         &              &               &                & 2019-Mar-17 & 1.10 & J0019+2021 & J0238+1636 & 1.9      \\
                                                      
         A68-HLS115     & 00:37:09.503 & +09:09:03.80   & 125.3--128.6 & 2018-Dec-16 & 0.95 & J0037+1109 & J0006$-$0623 & 4.9      \\
                        &              &                & 137.1--140.7 & 2018-Dec-17 & 0.95 & J0037+1109 & J0006$-$0623 & 5.1      \\
                  
         A68-C0         & 00:37:07.404 & +09:09:26.57  & 125.1--128.6 & 2018-Dec-17 & 0.95 & J0037+1109 & J0006$-$0623 & 4.2      \\
                        &              &                & 137.1--140.7 & 2018-Dec-18 & 0.95 & J0037+1109 & J0006$-$0623 & 4.3      \\
                                                     
         \hline
    \end{tabular}\\
\end{table*}\label{table: date}

\begin{table}
    \centering
    \caption{Noise levels in the maps and spectra resulting from our
        observations, where channel widths were binned to 15\,\kms.}
    \begin{tabular}{lcc}
    \hline \hline
    Target          &  Cont. noise level       & Spectral noise level \\
    galaxy          &  [mJy$\cdot$beam$^{-1}$] & [mJy$\cdot$beam$^{-1}$]        \\
    \hline    
    \target\        & 0.01                     & 0.01 \\
    MACS\,J0032-arc & 0.03                     & 0.08 \\
    A68-C0          & 0.02                     & 0.19 \\
    A68-HLS115      & 0.01                     & 0.22 \\
    \hline
    \end{tabular}
    \label{table: rms}
\end{table}

\section{Observational results}
\label{section: results}

\begin{figure*}
    \centering
        \includegraphics[width = \textwidth]{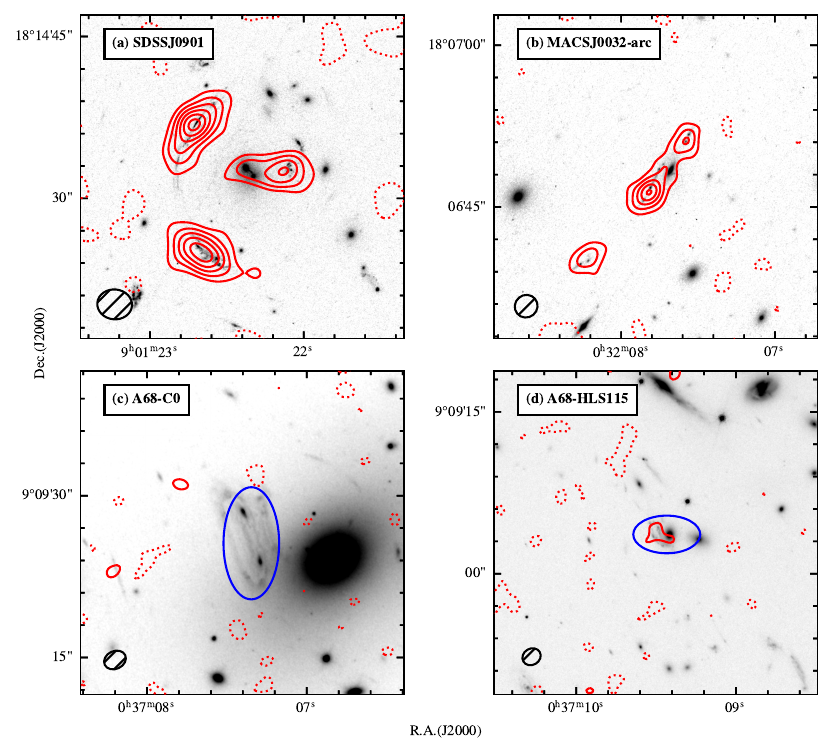}
        \caption{\label{fig: cont_fits} Millimetre continuum maps of the target
        galaxies. The top-left and top-right panels show {\it HST} IR images
        of \target\ and MACS\,J0032-arc, respectively; the bottom left and
        bottom right images are for the galaxies, A68-C0 and A68-HLS115,
        respectively. Red solid contours indicate detection boundaries with
        continuum signals 3, 5, 7, 9, 11 and 13 $\times$ the noise (as
        calculated in CASA); red dotted lines are contours of the zero level.
        Black ellipses in the bottom left corners are the synthesised beam
        sizes for the continuum maps and the blue ellipses in the middle of the
        lower panels are the regions of the targets, defined manually.  }
\end{figure*}

\begin{figure*}
    \centering
        \includegraphics*[width = \textwidth]{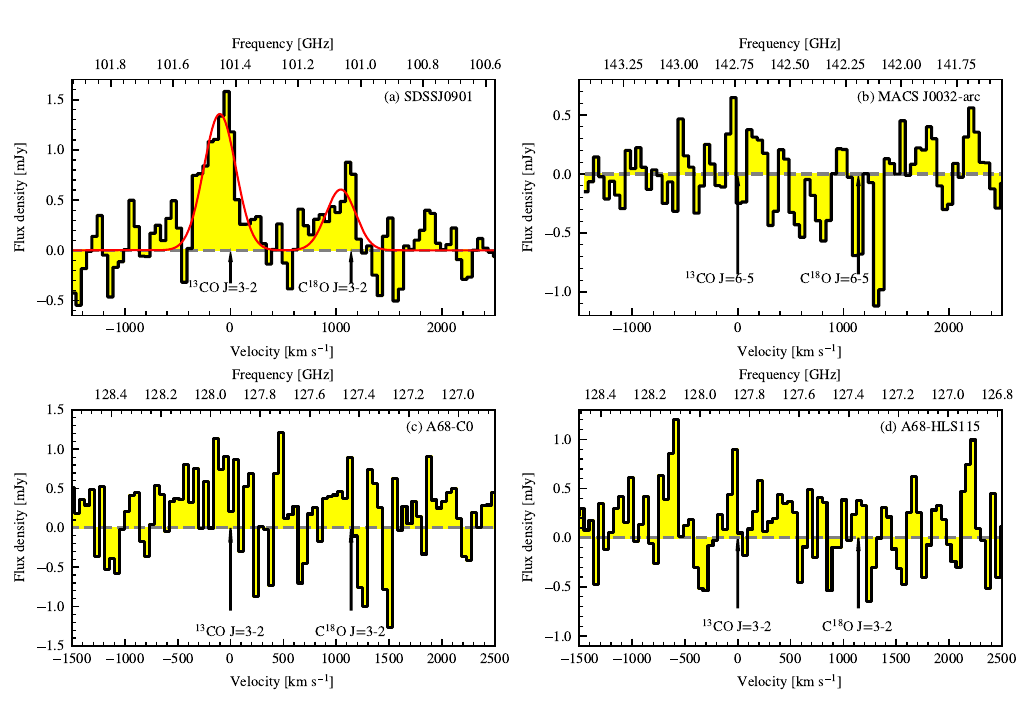}
        \caption{\label{fig: co_spectra}Spectra of \thco\ and \ceo\ towards our
        four targets. The velocity on the bottom horizontal axis is set to the
        velocity of \thco. The spectra have been smoothed to 50\,km\,s$^{-1}$.
        The grey dashed line denotes zero flux density; the red solid line is a
        double Gaussian fit to \thco\ and \ceo. Black arrows point to the
        central frequency of the \thco\ and \ceo\ lines.}
\end{figure*}

\begin{figure*}
    \centering
        \includegraphics[width = \textwidth]{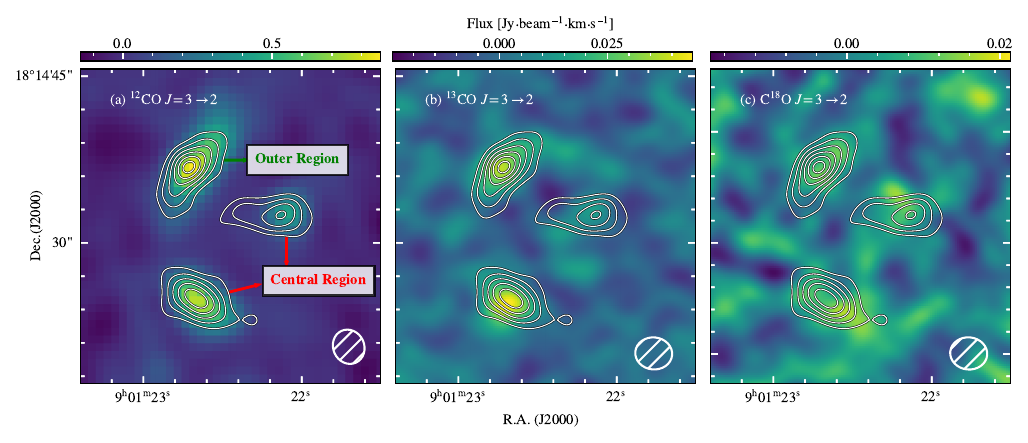}
        \caption{\label{fig: J0901_co} Images of \target. From left to right,
        the color maps are for the \twco\ $J$=3\to2, \thco\ $J$=3\to2 and \ceo\
        $J$=3\to2 emission lines, respectively.  White contours denote 3, 5, 7,
        9, 11, and 13-\sig\ continuum emission from our work(\sig\ = 0.01
        mJy$\cdot$beam$^{-1}$).  In the \twco map, the outer and central
        regions are labelled according to the lens modelling results in
        \citet{sharon2019}.}
\end{figure*}

\begin{figure}
    \centering
    \includegraphics*[width = \columnwidth]{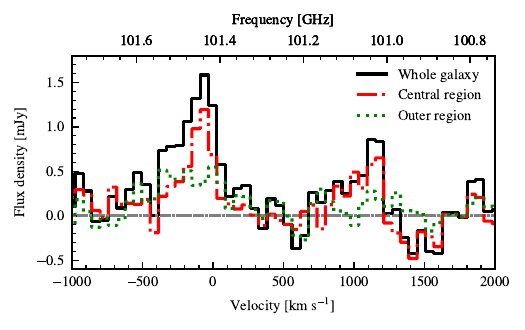}
    \caption{\label{fig: divide_spec} Spectra of \thco\ and \ceo\ in different
        parts of \target. The black-solid line represents the spectrum of the
        whole of \target. The red-dashed line represents the spectrum of the
        central region of the galaxy; the green-dotted line represents the
        spectrum of the outer regions of the galaxy.}
\end{figure}

\subsection{Submillimeter continuum}
\label{subsection: results-cont}

Continuum emission from \target\ and MACS\,J0032-arc was detected at the
$>$10-\sig\ level, while the signal-to-noise ratios for detections of continuum
towards A68-C0 and A68-HLS115 were below 5, as shown in Fig.~\ref{fig:
cont_fits}.

For the two galaxies with robust continuum detections, we used their 3-\sig\
continuum contours as the spatial distribution over which to extract their
spectrum. For the other two galaxies, we set the spatial range manually, as
shown in Fig.~\ref{fig: cont_fits} by the blue circles.

\subsection{Emission lines}
\label{subsection: results-line}

We have detected \thco\ and \ceo\ in \target, at 7.8~\sig\ and 3.3~\sig,
respectively. However, we did not detect any line emission at $>$ 3~\sig\
towards the other three targets. Fig.~\ref{fig: co_spectra} shows the spatially
integrated spectra of all four targets. The continuum contribution to the
spectra of \target\ and MACS\,J0032-arc have been subtracted. We did not
perform continuum subtraction for A68-C0 and A68-HLS115 because there is no
significant continuum emission (cf.\ Section~\ref{subsection: results-cont}).
For the spectrum of \target, we fitted Gaussian profiles to the line emission
(red line in Fig.~\ref{fig: co_spectra}). Fluxes and noise levels for the
detected lines and continuum -- along with the 3-\sig\ upper limits for the
non-detections -- are shown in Table~\ref{table: result}. The flux ratio of
\thco/\ceo\ in \target\ is $2.38 \pm 0.83$.

\begin{table*}
    \caption{\label{table: result} Observational results, for lines and continuum.}
    \centering
    \begin{tabular}{llc}
        \hline \hline
        \multicolumn{3}{c}{\textbf{Detected lines for \target}} \\
        Transition        & Flux                      & Line width (FWHM) \\
                          & [Jy\,km\,s$^{-1}$]        & [km\,s$^{-1}$] \\
        \hline
        \thco\ $J$=3\to2   & 0.47 $\pm$ 0.06             & 365 $\pm$ 54 \\
        \ceo\  $J$=3\to2   & 0.20 $\pm$ 0.06             & 214 $\pm$ 83 \\
        \twco\,$J$=3\to2   & 8.79 $\pm$ 0.61             & 344 $\pm$ 7 \\
        \hline
        \multicolumn{3}{c}{\textbf{Flux ratios of \target}} \\
        $I$(\thco)/$I$(\ceo)   & 2.38 $\pm$ 0.83          & \\
        $I$(\twco)/$I$(\thco)  & 18.7 $\pm$ 2.7         & \\ 
        \hline
        \multicolumn{3}{c}{\textbf{Upper limits of \thco\ and \ceo}} \\
        Target             & Transition & 3-\sig\ upper limit \\
        galaxy             & & [Jy\,km\,s$^{-1}$]  \\ 
        \hline 
        MACS\,J0032-arc     & 6--5 & <0.18 \\
        A68-C0             & 3--2 & <0.39 \\
        A68-HLS115         & 3--2 & <0.30 \\
    \hline
        \multicolumn{3}{c}{\textbf{Continuum flux density}} \\
        Target            & Flux density                & Frequency (obs.) \\
        galaxy            & [mJy]                        & [GHz]\\
        \hline
        \target           & 0.65 $\pm$ 0.16             & 106.40 \\ 
        MACS\,J0032-arc    & 0.82 $\pm$ 0.15             & 147.57 \\
        A68-C0            & 0.20 $\pm$ 0.16             & 132.91\\
        A68-HLS115        & 0.17 $\pm$ 0.07             & 132.96\\ 
        \hline \hline 
    \end{tabular}
\end{table*}

\subsection{Spatial distribution of isotopologues towards SDSS\,J0901+1814}
\label{subsection: results-spatial}

Fig.~\ref{fig: J0901_co} shows the spatial distribution of the $J$=3\to2
transitions of \twco, \thco\ and \ceo. All maps have similar beam sizes, shown
in the bottom-right corner of each sub-panel. The contours show the 3-mm
continuum emission, at the 3, 5, 7, 9, 11 and 13-\sig\ levels. For \twco, the
strongest of the three images is the north-eastern, while for \thco\ it is the
south-eastern. Similar to \twco, the western image of \thco\ does not show any
signal above 3~\sig. For \ceo, the emission from all of the three images is
weak. The \twco\ and \thco\ emission comes from well within the continuum
contours. Although the western image has a 9-\sig\ detection of its continuum,
all the line transitions show only weak emission there.

Due to the strong lensing effect \citep{kochanek2006}, the observed galaxy
images are highly distorted compared to that of the source plane.
\citet{sharon2019} performed a detailed lens modelling of \target, which
appears as a single galaxy in the source plane. The velocity structure of
\target\ resembles a regular, rotating disk with a radius of $\sim$8\,kpc,
disfavoring a galaxy merger scenario.

As shown by \cite{sharon2019}, the center of \target\ is transformed by lensing
into the south-eastern and western images. Accordingly, we classify these two
images as the central region, and the north-eastern image as the outer region.
We extracted two spectra (cf.\ Fig.~\ref{fig: divide_spec}) from the
north-eastern image, and the sum of the south-eastern and western images,
respectively, thus presenting spectra of the inner and outer parts of the
target galaxy. The \thco/\ceo\ ratios are $1.9\pm1.0$ and $2.9\pm1.2$ in the
central and outer regions of the target galaxy, \target, respectively. 

\newpage
\section{Galactic chemical evolution (GCE) model}
\label{section: GCEmodel}

Following the approach of \citet{romano2017,zhang2018} and \citet{romano2020},
we have adopted a GCE model to analyse our observational results. We employ the
open-source code, NuPyCEE \citep{cote2017}, to simulate the chemical evolution
of the targeted galaxies. Besides the observed isotopic abundance ratio, which
was the main focus of \citet{zhang2018}, here we could further compare
measurements of the stellar mass and oxygen abundances reported in the
literature \citep{sharon2019}. We built up chemical evolution tracks: a
one-zone model with several user-defined modifications. We focus on \target, in
which \thco\ and \ceo\ were detected. In the following, we list the key
parameters adopted in the model. 

\subsection{Stellar IMFs}
\label{subsection: GCE-IMF}

Several different IMFs were adopted in our GCE models. We modified the
original NuPyCEE code to better control the customised IMFs. 

The forms of the adopted IMFs were as follows:
\begin{equation}
        \xi(m) = \frac{dN}{dm} = \left\{
    \begin{array}{ll}
            k_1 \times m^{-\alpha_1}, & 0.1 \leq \frac{m}{M_\odot} < 0.5 ,\\
            k_2 \times m^{-\alpha_2}, & 0.5 \leq \frac{m}{M_\odot} < 1 ,\\ 
            k_2 \times m^{-\alpha_3}, & 1 \leq \frac{m}{M_\odot}< 100,
    \end{array}
    \right.
\end{equation}

\noindent
where $k_1$, $k_2$, and $k_3$ are parameters that impose continuity and
normalise the IMF to 1\,\Msun\ ; $m$ is the stellar mass. The power-law indices
of all four IMFs are listed in Table~\ref{table: imf}. 

In this work, the `Milky-Way IMF' is taken from the literature, namely
\cite{romano2017, zhang2018,romano2019}, which adopt the IMF slopes from
\citealt{1986FCPh...11....1S} and \citealt{kroupa1993}. The $\alpha_3$ slope,
in particular, which is found in star-forming regions in the Solar
neighbourhood \citep{kroupa1993}, best reproduces the isotopic ratios of the
Milky Way \citep{romano2017}. 

The other IMFs presented in Table~\ref{table: imf} are the same as those in
\citet{zhang2018}, except for the `top-heavy IMF', which is original to this
work, for which we have changed the $\alpha_3$ slope to 2.5. In this context,
`top-heavy' refers to a flatter $\alpha_3$ slope than that of the Milky-Way
IMF. 

\begin{table}
    \caption{Parameters of the IMFs used in this work.} \label{table: imf}
    \centering
    \begin{tabular}{ccccc}
        \hline \hline
        IMF model name & $\alpha_1$ & $\alpha_2$ & $\alpha_3$ & Reference \\
        \hline
        Bottom-heavy   & 2.7        & 2.7        & 2.7        & \citet{zhang2018} \\
        Milky-Way      & 1.3        & 2.2        & 2.7        & \citet{kroupa1993} \\
        Top-heavy      & 1.3        & 2.1        & 2.5        &
        -     \\
        Ballero        & 1.3        & 1.95       & 1.95       & \citet{ballero2007}\\
        \hline
    \end{tabular}
\end{table}

\subsection{Star-formation history}
\label{subsection: GCE-SFH}

The star-formation history (SFH) of \target\ is constrained by several
observational aspects, including the upper age limit that comes from its
spectroscopic redshift\footnote{The redshift of \target, $z=2.26$, corresponds
to a maximum evolutionary time of 2.9\,Gyr from the Big Bang.}, the various
multi-wavelength SFR estimates, and its stellar mass \citep{sharon2019}. 

The star formation is set to start 1.9\,Gyr after the Big Bang, and to last for
1\,Gyr. The SFR estimated from H$\alpha$ is 14.5\,\Msunyr, a value that is
likely to be heavily influenced by dust obscuration \citep{sharon2019}. We
therefore adopt the SFR of 268\,\Msunyr\ derived from the total IR luminosity
\citep[from 8 to 1000\,$\mu$m in the rest frame; ][]{sharon2019}. The two SFRs
were estimated assuming the IMF of \citet{kroupa2001}.

For all of our \target\ galaxy models, we adopt a stellar mass of $9.5 \times
10^{10}$\,\Msun, derived from \textit{Spitzer}/IRAC 3.6$\mu$m and 4.5 $\mu$m
imaging \citep{saintoge2013, sharon2019}. Given the same star-formation
duration of 1 Gyr, tracks with different IMFs lead to slightly different SFRs.
In particular, the more top-heavy the IMF, the larger the fraction of mass
locked in high-mass stars in each time interval. Given the short lifetimes of
massive stars (1--30 Myr), galaxies with a top-heavy IMF have lower stellar
masses after 1\,Gyr of evolution; thus, to form identical stellar masses at the
end of the computation, higher SFRs are required for models with more top-heavy
IMFs. 

We design four different histories to cover both secular evolution and the
most extreme starbursts: 

I) Flat SFH. In this case, the SFR has a fixed value during the 1-Gyr
star-forming period. The stellar mass and the chemical abundances of the galaxy
both increase smoothly. 

II) Exponentially declining SFH. This is inspired by the Kennicutt-Schmidt
star-formation relation \citep{kennicutt1998a} and galactic infall models
\citep{prantzos1998, spitoni2017}, and it is often adopted in cosmological
simulations. We specify a decreasing slope of -$10^{-9}$\,\Msun yr$^{-2}$ for
the curve, but relax the initial SFR to result in a final stellar mass of
$10^{11}$\,\Msun.

III) Exponentially increasing SFH. This resembles the typical SFH observed for
local dwarf galaxies \citep{deboer2012}. In this case, most elements/isotopes
are formed in the later evolutionary stages. 

IV) Recent intensive starburst SFH. This model simulates the most extreme
starburst galaxies, in which the burst started around 100\,Myr before the
observed epoch. Owing to the short timescale, stars with masses below
4--5\,\Msun\ do not contribute to the ISM enrichment, i.e.\ any \cth\
enrichment is limited to stars above 5\,\Msun.

The formation of secondary elements, such as \oei, relies highly on
pre-existing metals, so their abundances are sensitive to variations in the
early stages of star formation.

\subsection{Stellar yield tables}
\label{subsection: GCE-yield}

We have adopted the same stellar yield tables for low-metallicity AGB stars and
massive stars that were used in \citet{romano2017} and \citet{zhang2018}.
Crucially, these are capable of providing a satisfactory fit to the abundances
of the CNO isotopes in the Milky Way, i.e.\ those for which the most accurate
data exist \citep{romano2017}. For metal-rich AGB stars, we use the
state-of-the-art stellar yield table from \citet{cinquegrana2022}. To construct
the whole yields table, covering all metallicity and stellar mass ranges, we
use the same methods adopted in \citet{romano2017} and \citet{zhang2018}, where
some details are elaborated as follows: 

For massive stars, we adopt the supernova yields from \citet{nomoto2013} in the
mass range 13--40\,\Msun. For AGB stars with an initial mass of 1--6.5\,\Msun,
born with relatively low metallicities ($Z \leq 0.02$), we adopt the yields
from \citet{karakas2010}. For the stellar mass range, 6.5--8\,\Msun, at low
metallicities, the stellar yields are kept equal to those of 6.5-\Msun\ stars.
For AGB stars with initial masses of 1--8\,\Msun, born with relatively high
metallicities ($0.02 < Z < 0.05$), we adopt the yields from
\citet{cinquegrana2022}. In the gap between 8 and 13\,\Msun, we interpolate the
stellar yields linearly on the logarithmic scale in stellar mass and
metallicity. Lastly, for the mass range, 40--100\,\Msun, we adopt the stellar
yields of 40\,\Msun\ stars, the most massive in \citet{nomoto2013}. Type~Ia
supernovae (SN) are not considered because the yield contributions to \cth\ and
\oei\ in Type~Ia SN are below $10^{-5}\times$ those of their main isotopes
\citep{seitenzahl2013}.

\subsection{IMF-weighted stellar yields}
\label{subsection: results-yield}

\begin{figure}
    \centering
        \includegraphics[width = \columnwidth]{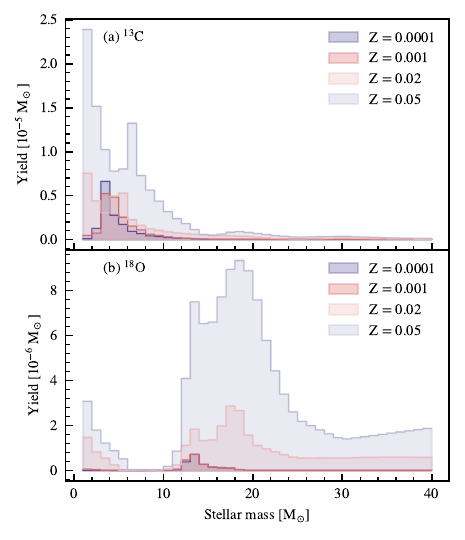}
        \caption{\label{fig: yield} IMF-weighted yields (weighted by the
        Milky-Way IMF) for \cth\ and \oei\ in single stellar populations with
        different metallicities, represented by different colors. Contribution
        ratios for different types of stars are summarised in Table~\ref{table:
        yield}. At metallicities of 0.001 and 0.0001, the stellar yields of \oei\
        show only very small differences across the stellar mass range.}
\end{figure}

\begin{table}
        \caption{\label{table: yield} Production ratio of isotopes from
        Low-and-Intermediate Mass Stars (LIMS, M $< 8$\,\Msun) and massive
        stars ($M \geq 8$\,\Msun) for a single stellar population following the
        Milky-way IMF.  The mass of the stellar population is normalized to
        1\,\Msun. The IMF adopted for weighting is the Milky-Way IMF shown in
        Table \ref{table: imf}. Here, metallicity (Z), presents the mass
        fraction of the elements heavier than hydrogen and helium. } \centering
        \footnotesize
        \begin{tabular}{c c c c c}
                \hline \hline
                Nuclei & Z           & LIMS (\%) & Massive stars (\%) & Total yield (\Msun) \\
                \hline
                \cth\  & 0.0001      & 92.9               & 7.1     & 2.4 $\times 10^{-5}$         \\
                       & 0.001       & 92.0               & 8.0     & 2.8 $\times 10^{-5}$         \\
                       & 0.02        & 84.8               & 15.2     & 6.1 $\times 10^{-5}$         \\
                       & 0.05        & 84.4               & 15.6     & 1.8 $\times 10^{-4}$         \\
                \hline
                \oei\  & 0.0001      & 1.0                & 99.0     & 1.8 $\times 10^{-6}$         \\
                       & 0.001       & 10.1               & 89.9     & 2.1 $\times 10^{-6}$         \\
                       & 0.02        & 21.9               & 78.1     & 2.9 $\times 10^{-5}$         \\
                       & 0.05        & 14.1               & 85.9     & 2.1 $\times 10^{-4}$         \\
                \hline
        \end{tabular}
\end{table}

Fig.~\ref{fig: yield} shows the adopted \cth\ and \oei\ stellar yields
\citep{karakas2010, nomoto2013, cinquegrana2022} of a single stellar population
(from Section~\ref{subsection: GCE-yield}), weighted by the Milky-Way IMF
\citep{kroupa1993} . For comparison, the masses of the stellar populations are
all normalised to 1\,\Msun. Fig.~\ref{fig: yield} shows that most of the \cth\
is produced by LIMS while most of the \oei\ is produced by massive stars. The
specific production ratios also depend on the metallicity, $Z$, where we choose
four metallicities to calculate the ratios, as shown in Table~\ref{table:
yield}. 

For all metallicities, over 80\% of \cth\ comes from LIMS, and over 75\% of
\oei\ comes from massive stars. This strong differential explains why the mass
ratio of \cth/\oei\ is so sensitive to the IMF. The total production masses of
\cth\ and \oei\ both increase with metallicity. The production of \oei, as a
purely secondary element, is more sensitive to metallicity, compared with that
of \cth\, as mentioned later in Section~\ref{subsection: discussion-SFH}.
Single stellar populations produce a higher proportion of \oei\ with a
top-heavy IMF than with the Milky-Way IMF. The major origin of \cth\ is stars
at around 3\,\Msun, whose lifetimes are around 200\,Myr, while for \oei\ it is
18-\Msun\ stars with lifetimes of around 10\,Myr.

\subsection{Stellar mass evolution}
\label{subsection: GCE-Mstar}

All models have the same mass of living stars at the epoch that we have
observed in the targeted galaxy, i.e.\ at $z=2.26$ for \target. However,
NuPyCEE was not able to calculate the mass of the living stars. Therefore, we
make a user-defined model to trace the stellar mass evolution, namely,
\textit{stellar\_evol} \citep{stellar_evol2024}. This code is available in the
Zenodo repository: doi:10.5281/zenodo.11118895, and is also available on
github\footnote{\url{https://github.com/GuoZiYi-astro/stellar_evol}}. This
model considers the IMF and stellar lifetimes, and it enables the derivation of
the mass of living stars at any given evolutionary epoch.

\subsection{Miscellaneous}
\label{subsection: GCE-miscellaneous}

Our stellar lifetime table was taken from \citet{schaller1992}, to be
consistent with the stellar lifetimes implemented in the models by
\citet{romano2017} which are used in \citet{zhang2018}. In this table, the
shortest lifetime is around 4\,Myr for 50-\Msun\ stars. The default evolution
in NuPyCEE is processed on a logarithmic scale, which cannot precisely model
the chemical evolution at the later stages of the galaxy evolution. Therefore,
we track the evolution on a linear scale with a fixed timestep of 4\,Myr. This
ensures that newly formed, massive stars and their returned stellar yields are
accounted for properly.

Compared to the modelling work in \citet{zhang2018}, our work adopts a
different GCE framework, NuPyCEE, slightly different stellar yield tables, a
fixed evolutionary timestep setting, and without using the
Q-matrix\citep{talbot1973}. However, by adopting the same SFH and IMF as
\citet{zhang2018}, we can obtain the same ratio of \cth/\oei\ and almost
identical evolutionary tracks. This is re-assuring in terms of the robustness
of our user-defined code modification. 


\section{Comparison of observations with GCE models}
\label{section: comparison}

\begin{figure} 
\centering 
\includegraphics[width=0.95\columnwidth]{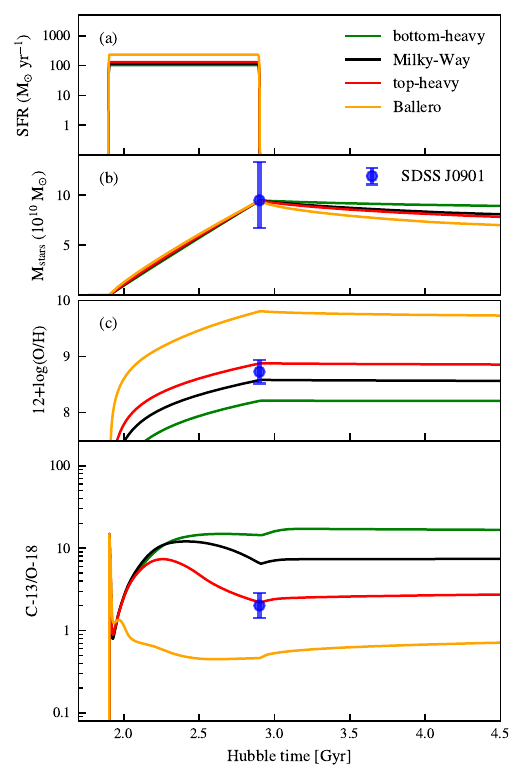}
\caption{\label{fig: co_fitting}
        Predicted evolutionary tracks of the \cth/\oei\ abundance ratio for galaxies
        with different IMFs, produced using NuPyCEE. Green curves are the GCE
        results assuming the bottom-heavy IMF; black curves assume the
        Milky-Way IMF \citep{kroupa1993}; red curves assume the top-heavy IMF;
        yellow curves assume the Ballero IMF. Blue circles are the observed
        stellar mass and oxygen abundance in \target, adopted from
        \citet{sharon2019}. The parameters of all the IMFs are listed in
        Table~\ref{table: imf} and the SFR is IMF-dependent.} 
\end{figure}

Fig.~\ref{fig: co_fitting} presents a comparison between our observed line
ratio, from Section~\ref{section: results}, and the GCE model predictions, as
described in Section~\ref{section: GCEmodel}. The curves show the evolutionary
tracks of SFR, stellar mass, oxygen abundance and \cth/\oei\ ratios for models
corresponding to our different adopted IMFs, from bottom-heavy to top-heavy.
All SFR tracks have the same slope, the same star-formation duration, and the
same final stellar mass, as mentioned in Section~\ref{subsection: GCE-SFH}. 

The stellar masses of all models reach $9.5 \times 10^{10}$\,\Msun\ after
3\,Gyr, as we required. After the cessation of star formation, the stellar
masses start to decrease slowly. The stellar mass of the model with the most
top-heavy IMF falls fastest, because it has the highest fraction of short-lived
massive stars.

The oxygen abundances of all models increase rapidly after the star formation
starts and continues to increase monotonically until the star formation ceases.
After that, the oxygen abundance of all models remains stable. The models with
the most top-heavy IMFs tend to end with higher oxygen abundance.

During the first 50--100 Myr of star formation, the \cth/\oei\ abundance ratio
increases first, and then decreases rapidly, for all models. After the lowest
dip, the isotopic ratios of all models, except the one with the Ballero IMF,
increase slowly to its second peak value and then decrease gradually. The
Ballero-IMF model has a small bump and decreases rapidly until star formation
ceases.

The \cth/\oei\ ratio agrees with the model that assumes a top-heavy IMF; it
cannot be explained by the other three IMFs (including the Milky-Way IMF).
Under the Milky-Way IMF, our \cth/\oei\ evolutionary track is roughly
consistent with that in \citet{zhang2018}, which is optimised for starburst
galaxies. The observed oxygen abundance (12+log(O/H)) can match both the Kroupa
and the top-heavy IMFs; the other two IMFs do not match.

\begin{figure} 
\centering 
\includegraphics[width=0.95\columnwidth]{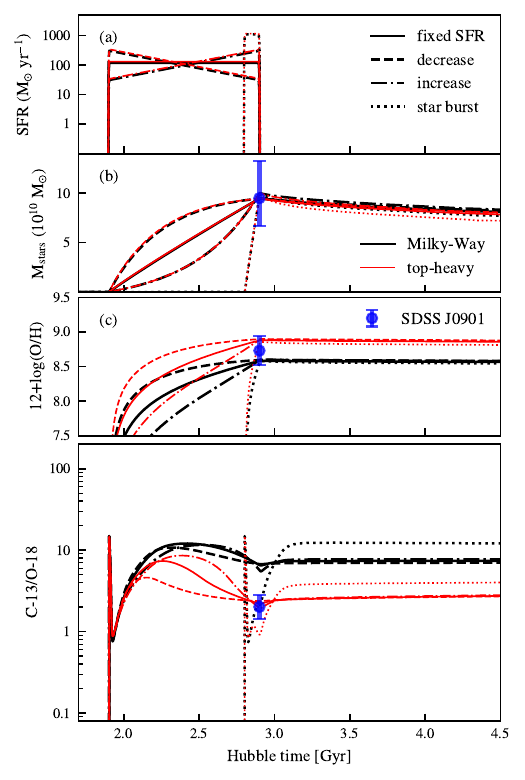}
\caption{\label{fig: diff_sfh}Theoretical predictions for the \cth/\oei\
        abundance ratio: evolutionary tracks of galaxies with different SFHs
        are shown by different line styles. Solid lines are the constant SFR
        model, which is the same as the SFH of Fig.~\ref{fig: co_fitting}.
        Dashed lines are the decreasing SFH; dash-dotted lines are the
        increasing SFH; dotted lines are the intense starburst SFH. Different
        line styles represent different IMFs implied in the models. Black
        curves represent the Milky-Way IMF \citep{kroupa1993}; red curves
        represent the top-heavy IMF (defined in Table~\ref{table: imf}). Blue
        dots represent our observations of \target.} 
\end{figure}

As described in Section~\ref{subsection: GCE-SFH}, we assume four different
SFHs, aiming to break the degeneracy between the SFH and IMF. The oxygen
abundance can be explained by both the Milky-Way and the top-heavy IMFs,
leaving some ambiguity, so it is necessary to test whether an extreme SFH with
a Milky-Way IMF could achieve the observed \cth/\oei\ ratio.

We investigated eight models, which consist of two different IMFs -- the
Milky-Way and top-heavy IMFs, and four different SFHs, to trace the evolution
of SFR, stellar mass, oxygen abundance and \cth/\oei\ ratios. These models are
presented in Fig.~\ref{fig: diff_sfh}. 

In Fig.~\ref{fig: diff_sfh}, all model tracks show strong variations due to the
IMF, no matter which SFH is adopted. Although different SFHs would introduce
individual biased evolutionary tracks, all models are designed to reproduce the
same stellar mass at the epoch when we observe \target. Varying the SFH does
not have a strong impact on the oxygen abundances, which are more sensitive to
the choice of IMF; all predicted oxygen abundances are within 1-$\sigma$ of the
observational data for \target.

Measurements of the stellar mass and the oxygen abundance help to reinforce the
constraints from the GCE modelling, compared to the dusty starburst galaxies
observed in \citet{zhang2018}. Taken alone, the oxygen abundance is not enough
to discriminate between the IMFs, but only the models with a top-heavy IMF can
explain the observed \cth/\oei\ ratio. No matter which SFH is applied in the
model, the observed \cth/\oei\ value cannot be reproduced with the Milky-Way
IMF within the observational uncertainties.


\section{Discussion}
\label{section: discussion}

\subsection{The influence of the star-formation history}
\label{subsection: discussion-SFH}

As shown in Fig.~\ref{fig: diff_sfh}, the SFH only slightly influences the
evolution of \cth/\oei, for the following reasons:

First, the ratio between the stellar yield of \cth\ and \oei\ relies strongly
on the initial metallicity of the stars \citep[e.g.][]{wiescher2010}. As
demonstrated in Section~\ref{subsection: results-yield} and Fig.~\ref{fig:
yield}, contribution ratios of \cth\ and \oei\ all rely on metallicity.
Compared to \cth, the stellar yield of \oei\ is more sensitive to metallicity,
because \cth\ is both a primary and a secondary element, which can be produced
by the first generation of stars. On the other hand, \oei\ is a pure secondary
element and can only be produced by stars containing $^{14}$N seeds at the
moment the star was born \citep{wilson1992, romano2017}. Most \oei\ is
therefore produced in the late, metal-rich stage of galaxy evolution. It is
this timing discrepancy which leads to the differences in the chemical
enrichment of \cth\ and \oei.

For all models in Figs.~\ref{fig: co_fitting} and \ref{fig: diff_sfh}, the
\cth/\oei\ abundance ratios start from a high level in the earliest few Myrs,
quickly drop, then present an increasing trend. In this phase, a low \cth/\oei\
ratio can persist for $\lesssim$ 100 Myr, during which the galaxy is still very
metal-poor and most \cth\ is not yet released to the ISM. At this early
evolutionary stage, both stellar mass and metallicity are orders of magnitude
lower than the final, observed values. 

Second, the release times of \cth\ and \oei\ are different. \cth\ and \oei\ are
mostly produced by LIMS and massive stars, which have longer and shorter
stellar lifetimes \citep{schaller1992}, respectively. After the initial drop
(see Fig.~\ref{fig: co_fitting} (d)), the abundance ratio of \cth/\oei\
increases slowly as the galaxy evolves and the \cth\ is released gradually by
LIMS. 

Third, if the star-formation timescale is longer than 2\,Gyr, the contribution
of novae becomes non-negligible \citep{dantona1982,romano1999,romano2003}.
However, nova outbursts mainly enrich isotopic abundances of \cth, $^{15}$N and
$^{17}$O \citep{starrfield1972, starrfield1974, romano2003}, compared to that
of \oei. Including the contribution of novae would thus reinforce our
conclusion that the IMF must be top-heavy, since it would increase the
\cth/\oei\ ratio in the models. 

With a Milky-Way IMF, only if such a massive galaxy is formed over a timescale
shorter than 100\,Myr, where the corresponding SFR is above 1000\,\Msunyr,
could the \cth/\oei\ ratio match the observed value. This is demonstrated by
the starburst model in Fig.~\ref{fig: diff_sfh}, in which a flat burst history
with an SFR of 1000\,\Msunyr\ results in a \cth/\oei\ ratio of $\sim 2$, below
the other SFH tracks with the Milky-Way IMF. 

With the same IMF, a single starburst model would predict a slightly higher
(within a factor of $2\times$) \cth/\oei\ ratio at the evolutionary end point,
compared to those produced with other more slowly-evolving SFHs, as shown in
Fig.~\ref{fig: diff_sfh}. This is because most stars were formed quickly in
low-metallicity conditions. Given the relatively low production of \oei\ from
massive stars at low metallicities, as shown in Fig.~\ref{fig: yield}, the
final \cth/\oei\ ratio could be slightly increased. This small difference would
not affect our main results. 

\subsection{Spatial distribution of the IMF}
\label{subsection: discussion-spatial}

Aided by strong lensing, we can partially resolve the spatial distribution of
\target. As shown in Section~\ref{subsection: results-spatial}, \thco/\ceo\
isotopologue flux ratios show differences amongst the three lensed images. The
line ratio from the outer region is slightly higher than that of the galactic
center. This indicates that the IMF in the central region might be more
top-heavy than that in the outer region. 

If confirmed, this phenomenon would be in line with the integrated galactic IMF
(IGIMF) theory \citep{kroupa2003, weidner2005, kroupa2013}. In that theory,
both the slope of the mass function of the embedded clusters and the IMF rely
on the SFR \citep{yan2017}. In \citet{sharon2019}, the SFR of \target\ shows a
decreasing radial distribution, consistent with spiral galaxies in the local
Universe \citep{gonzalez2016}, in both the image and the source plane. Since
the SFR density is highest in the galaxy center, the final integrated IMF would
also be most top-heavy in that same central region. However, the sensitivity of
the current data is somewhat limited, and we merely regard the observed spatial
variation of the IMF in \target\ as a plausible scenario, in need of
confirmation. 

\subsection{Impact on the cosmic evolution of star formation}
\label{subsection: discussion-SFRcosmic}

Variations of the IMF can affect measurements of the SFR, because most SFR
tracers -- ultraviolet radiation, H$\alpha$ emission, total infrared
luminosity, etc.\ -- are linked to the effects of massive, young stars
\citep[e.g.][]{kennicutt1998a}. If the IMF in a galaxy deviates from that seen
in the Milky Way, which has been used to benchmark most SFR tracers
\citep{kennicutt1998a}, then the SFR deduced for that galaxy will be
systematically in error. Our work implies that the IMF in high-redshift
main-sequence galaxies, which are believed to dominate the SFR density at and
around `cosmic noon' ($z\approx1$--3), is more top-heavy than the IMF of the
Milky Way. These galaxies thus contain a higher fraction of massive stars than
the Milky Way, and therefore produce a higher luminosity per unit stellar mass
in ultraviolet radiation, H$\alpha$ emission, total infrared luminosity, etc.,
meaning that their real SFRs (and total stellar masses) are likely rather lower
than those calculated in the literature to date \citep{jerabkova2018}.

With \textit{stellar\_evol}\ we estimate the proportion of massive stars among
all living stars in our simulations, for different assumed IMFs. For the same
total stellar mass, the fraction by mass of massive stars for the Milky-Way IMF
and the top-heavy IMF are 1.5\textperthousand\ and 2.5\textperthousand,
respectively. Thus if main-sequence galaxies at high redshift -- as represented
by \target\ -- have top-heavy IMFs, then their SFR estimates could decrease by
more than 40\%, consistent with recent findings using the {\it James Webb Space
Telescope} in high-redshift galaxies \citep[][cf.\
\citealt{cullen2023}]{inayoshi2022, wang2023}.

More detailed calculations and benchmarking for SFR tracers are beyond the
scope of this work. 

\section{Summary}

In this paper, we report observations of CO isotopologues in four strongly
lensed main-sequence galaxies at redshift $\sim 2$. A robust detection of
\thco\ and \ceo\ emission is obtained from one target, \target, at $z \sim
2.26$. This is the first detection of CO isotopologues in a main-sequence
galaxy at high redshift. The observed flux ratio of \thco/\ceo\ in this galaxy
is $2.4 \pm 0.8$, which is lower than the average value observed for local
main-sequence galaxies. Via detailed galactic chemical evolution modelling, we
find that only an IMF more top-heavy than the Milky Way IMF can reproduce the
observed isotopic ratio, oxygen abundance, and stellar mass, simultaneously.
The power-law indices of Milky Way IMF and the top-heavy IMF differ by only a
small amount of 0.2, which already changes the \cth/\oei\ ratio by a factor of
3--4$\times$, which showcases the sensitivity of the isotopic method. Our
result suggests that main-sequence galaxies in the high-redshift Universe may
prefer a top-heavy IMF, and therefore, create more massive stars than
previously thought.


\begin{acknowledgments}

We thank Donatella Romano and Francesca Matteucci for very helpful discussions
and extensive support. We acknowledge the support of the National Natural
Science Foundation of China (NSFC) under grants 12173016, 12041305, 12203100,
12203021. We acknowledge the Program for Innovative Talents, Entrepreneur in
Jiangsu. We acknowledge the science research grants from the China Manned Space
Project, CMS-CSST-2021-A08 and CMS-CSST-2021-A07. Z.Q.Y.\ acknowledges the
support from the Jiangsu Funding Program for Excellent Postdoctoral Talent under
grant number 2022ZB54, and the Fundamental Research Funds for the Central
Universities under grant number 0201/14380049. AWSM acknowledges the support of
the Natural Sciences and Engineering Research Council of Canada (NSERC) through
grant reference number RGPIN-2021-03046.

This paper makes use of the following ALMA data: ADS/JAO.ALMA\#2016.1.00406.S,
ADS/JAO.ALMA\#2018.1.00588.S. ALMA is a partnership of ESO (representing
its member states), NSF (USA) and NINS (Japan), together with NRC (Canada),
MOST and ASIAA (Taiwan), and KASI (Republic of Korea), in cooperation with the
Republic of Chile. The Joint ALMA Observatory is operated by ESO, AUI/NRAO and
NAOJ.

We obtained {\it HST} WFC1 and WFC3 images from the MAST archive at STScI. 

\end{acknowledgments}


\appendix

\bibliographystyle{aasjournal}
\bibliography{main}{}

\end{document}